\documentclass[reviewcopy]{elsart}

\usepackage{graphicx}
\usepackage{subfigure}
\usepackage{amsmath}

\journal{Optics Communications}
\newcommand{\avg}[1]{\left\langle #1 \right\rangle}
\renewcommand{\Im}{\text{Im}\;}
\renewcommand{\Re}{\text{Re}\;}

\begin{document}

\begin{frontmatter}

\title{Phase control algorithms for focusing light through turbid media}
\author{I.M. Vellekoop\corauthref{cor1}},
\corauth[cor1]{Corresponding author.} \ead{i.m.vellekoop@utwente.nl}\ead[url]{cops.tnw.utwente.nl}
\author{A.P. Mosk}

\address{Complex Photonic Systems, Faculty of Science and Technology, and MESA$^+$ Institute for
Nanotechnology, University of Twente, P.O.Box 217, 7500 AE Enschede, The Netherlands}

\date{}

\begin{abstract}
Light propagation in materials with microscopic inhomogeneities is affected by scattering. In scattering
materials, such as powders, disordered metamaterials or biological tissue, multiple scattering on
sub-wavelength particles makes light diffuse. Recently, we showed that it is possible to construct a
wavefront that focuses through a solid, strongly scattering object. The focusing wavefront uniquely
matches a certain configuration of the particles in the medium. To focus light through a turbid liquid or
living tissue, it is necessary to dynamically adjust the wavefront as the particles in the medium move.
Here we present three algorithms for constructing a wavefront that focuses through a scattering medium. We
analyze the dynamic behavior of these algorithms and compare their sensitivity to measurement noise. The
algorithms are compared both experimentally and using numerical simulations. The results are in good
agreement with an intuitive model, which may be used to develop dynamic diffusion compensators with
applications in, for example, light delivery in human tissue.
\end{abstract}

\end{frontmatter}

\maketitle
\section{Introduction}
\begin{figure}
  \includegraphics[width=7.5cm]{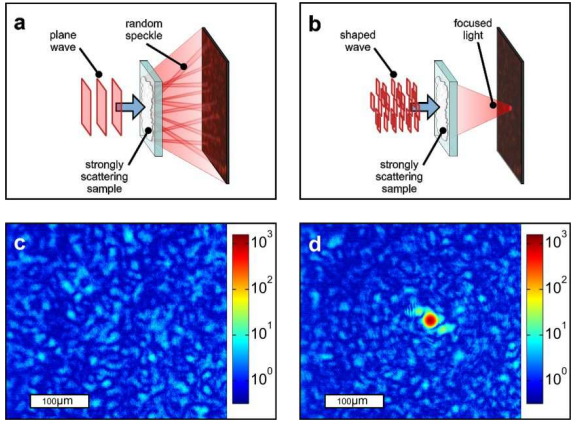}\\
  \caption{Principle and experimental results of inverse wave diffusion. a) A multiply scattering object
  destroys the spatial coherence of incident light. b) When the same object is illuminated
  with a specially constructed matching wavefront, the transmitted light focuses to a tight spot. c) Recorded intensity
  transmission of an unshaped wave through a 10 $\mu$m thick layer of TiO$_2$ pigment. d) Intensity
  transmission through the same sample with a shaped wavefront.}\label{fig:inverse-diffusion}
\end{figure}

\noindent Materials such as paper, white paint or human tissue are non-transparent because of multiple
scattering of light \cite{Milne1921,Chandrasekhar1960,Ishimaru1978}. Light propagating in such materials
is diffuse. Recently, we have shown that coherent light can be focused through diffusive media yielding a
sharp, intense focus \cite{Vellekoop2007}. Starting with the situation where a scattering object (a layer
of TiO$_2$ pigment with a thickness of approximately 20 transport mean free paths) completely destroys the
spatial coherence of the incident light (Fig. \ref{fig:inverse-diffusion}a, \ref{fig:inverse-diffusion}c),
we controlled the incident wavefront to exactly match scattering in the sample. Afterwards, the
transmitted light converged to a tight, high contrast focus (Fig. \ref{fig:inverse-diffusion}b,
\ref{fig:inverse-diffusion}d). These matched wavefronts experience inverse diffusion, that is, they gain
spatial coherence by travelling through a disordered medium.

For a given sample of scattering material, there is a unique incident wavefront that makes the object
optimally focus light to a given point. Like a speckle pattern, this wavefront is disordered on the scale
of the wavelength of light. This wavefront cannot be constructed from a small number of smooth base
functions, which unfortunately renders the efficient algorithms used in adaptive optics (see e.g.
\cite{Tyson1998}) ineffective. In Ref. \cite{Vellekoop2007}, we presented an algorithm that finds the
optimal wavefront when the sample is perfectly stationary and the noise level is negligible. To find
applications in, for example, fluorescence excitation or photodynamic therapy, the wavefront has to be
adjusted dynamically as the scatterers in the sample move. In this paper, we present two additional
algorithms, that dynamically adjust the wavefront to follow changes in the sample. The performance of the
algorithms is in good agreement with numerical simulations and with an analytical model. We show that the
new algorithms are superior to the original algorithm when the scatterers in the sample move or when the
initial signal to noise ratio is poor.

Wave diffusion is a widely encountered physical phenomenon. The use of multiple scattered waves is the
subject of intensive study in the fields of, for instance, ultrasound imaging
\cite{Fink1999,Lobkis2001,Borcea2005}, radio and microwave antennas \cite{Lerosey2007,Foschini1996},
seismography \cite{Shapiro2005}, submarine communication \cite{Kuperman1998}, and surface plasmons
\cite{Stockman2002}. While the algorithms discussed in this paper were developed for spatial phase shaping
of light, they can be used for any type of wave and apply to spatial phase shaping as well as to frequency
domain phase shaping (also known as coherent control, see e.g. \cite{Herek2002}) as the concepts are the
same.

This article is organized as follows. First the key concepts of inverse diffusion are introduced and the
three different algorithms are presented. Then the experimental apparatus is explained and the measured
typical performance of the algorithms is compared. In the subsequent section, we will compare the
experimental results with numerical simulations and analyze the data in terms of noise and stability of
the scatterers. Finally, we will analytically explain the characteristic features of the different
algorithms and discuss their sensitivity to noise.

\section{Algorithms for inverse diffusion}\label{sec:algorithms}
\begin{figure}
  \includegraphics[width=7.5cm]{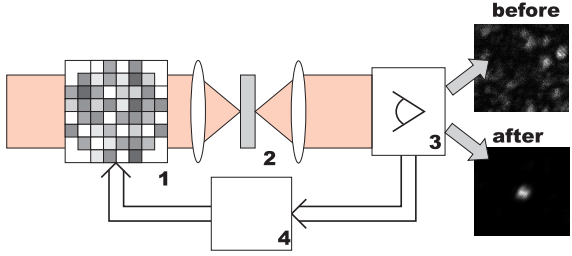}\\
  \caption{Feedback loop for achieving inverse diffusion. An incident monochromatic beam is shaped using a spatial light modulator (1)
  and projected on a non-transparent multiply scattering object (2). A detector (3) detects the amount of transmitted light that reaches
  the target area. A feedback algorithm (4) uses the signal from the detector to program the phase modulator. Before the algorithm is
  started, the transmitted light forms a random speckle pattern. The algorithm changes the incident wave
  to increase the intensity in the target area. After a few iterations, the transmitted light focuses on the target.}\label{fig:block-diagram}
\end{figure}

\noindent The key elements of an inverse diffusion setup are a multiply scattering sample, a spatial light
modulator, an optimization algorithm and a detector, as shown in Fig. \ref{fig:block-diagram}. The sample
can be anything that scatters light without absorbing it. We will consider only samples that are thicker
than approximately 6 transport mean free paths for light. Light transmitted through these samples is
completely diffuse and the transmitted wavefront is completely scrambled, i.e., it has no correlation with
the incident wavefront \cite{Pappu2002}.

The incident wavefront is constructed using a spatial phase modulator. The modulator consists of a
2D-array of pixels that are grouped into $N$ equally sized square segments. A computer sets the phase
retardation for each of the segments individually to a value between $0$ and $2\pi$. The optimization
algorithm programs the phase modulator based on the detector output. Since the sample completely scrambles
the incident wavefront, all segments of the wavefront are scattered independently and the optimal
wavefront will not be smooth.

Behind the sample is a detector that provides feedback for the
algorithm. The detector defines the target area where the intensity
is maximized. The field at the detector is the result of
interference from scattered light originating from the different
segments of incident wavefront. When the phase of one or more
segments is changed, the target intensity responds sinusoidally. We
sample the sine wave by taking 10 measurements. The process of
capturing a single sine wave and possibly adjusting the phase
modulator accordingly is called an iteration.

The amount of control we have over the propagation of light in the disordered system is quantified by the
signal enhancement. The enhancement $\eta$ is defined as
\begin{equation}
\eta \equiv \frac{I_N}{\avg{I_0}},
\end{equation}
where $I_N$ is the intensity in the target after optimization and $\avg{I_0}$ is the ensemble averaged
transmitted intensity before optimization. In a perfectly stable system, the enhancement is proportional
to $N$ \cite{Vellekoop2007}, meaning that the more individual segments are used to shape the incident
wavefront, the more light is directed to the target. In practice, however, the enhancement is limited by
the number of iterations that can be performed before the sample changes too much. We define the
persistence time $T_p$ as the decay time of the field autocorrelate of the transmitted speckle, which is a
measure for the temporal stability of the sample. The persistence time depends on the type of sample and
on environmental conditions. Typical values of $T_p$ range from a few milliseconds in living tissue
\cite{Briers1995} to hours for solid samples in laboratory conditions. The other relevant timescale is the
time required for performing a single iteration, $T_i$. In our experiments, we operate the phase modulator
at just below 10 Hz and take ten measurements for each iteration; we have $T_i\approx 1.2$s.

We will now present three algorithms we used to invert wave
diffusion. The advantages and disadvantages of the algorithms are
discussed briefly and will be analyzed in detail later.

\begin{figure}
  \includegraphics[width=7.5cm]{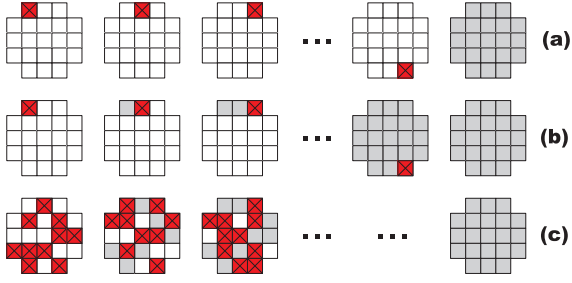}\\
  \caption{Principle used in the three different optimization algorithms. a) For the stepwise sequential algorithm, all segments are addressed sequentially
  (marked squares). After the optimal phase is measured for all segments, the modulator is updated to construct the optimal
  wavefront (light gray squares). b) The continuous sequential algorithm is equal to the first algorithm, except that the modulator is updated after each iteration.
  c) The partitioning algorithm randomly selects half of the segments and adjusts their overall phase. The modulator is updated after each measurement. }\label{fig:algorithms-explained}
\end{figure}

\subsubsection{The stepwise sequential algorithm}
\noindent The stepwise sequential algorithm that was used in Ref. \cite{Vellekoop2007} is very
straightforward. The computer consecutively cycles the phase of each of the $N$ segments from 0 to $2\pi$.
The feedback signal is monitored and the phase for which the target intensity is maximal is stored. After
all iterations are performed, the phase of each segment is set to this optimal value (see Fig.
\ref{fig:algorithms-explained}a). In absence of measurement noise or temporal instability, algorithm 1 is
guaranteed to find the global maximum in the lowest number of iterations possible. However, when $N T_i\gg
T_p$, the speckle pattern decorrelates before all measurements are performed and the algorithm will not
work. Therefore, it is important to adjust the number of segments to the persistence time.

\subsubsection{The continuous sequential algorithm}
\noindent The continuous sequential algorithm is very similar to the stepwise sequential algorithm except
for the fact that the phase of each segment is set to its maximum value directly after each measurement
(see Fig. \ref{fig:algorithms-explained}b). This approach has two advantages. First of all, the algorithm
runs continuously and dynamically follows changes in the sample's scattering behavior. Furthermore, the
target signal starts to increase directly, which increases the signal to noise ratio of successive
measurements. It is still necessary to adjust $N$ to the persistence time $T_p$.

\subsubsection{The partitioning algorithm}
\noindent As an alternative to the two sequential algorithms, we propose a partitioning algorithm that
requires no a-priori information about the sample's stability. Each iteration the phase modulator is
divided randomly into two subsets, both containing half of the segments (Fig.
\ref{fig:algorithms-explained}c). The target intensity is maximized by changing the phase of one subset
with respect to the other. Since the phase of half of the segments is changed, the initial increase in
intensity will be fast and the feedback signal will be maximal. Therefore, this algorithm is expected to
be less sensitive to noise and to recover from disturbances more rapidly.

\section{Experiment}
\begin{figure}
  \includegraphics[width=7.5cm]{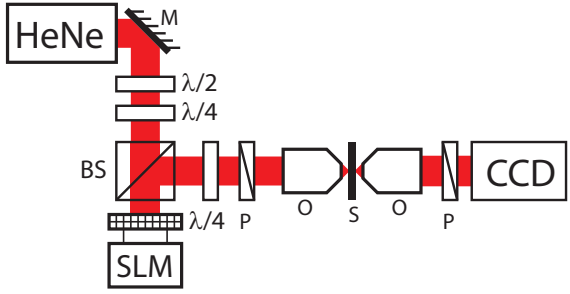}\\
  \caption{Experimental apparatus used for inverting diffusion. Light from a HeNe laser is spatially modulated by
  a liquid crystal spatial light modulator (SLM). Wave plates and a polarizer are used to generate and select the polarization
  state for which the modulator works in phase mostly mode. The shaped beam is focused on the sample. A reference
  detector monitors the total intensity falling on the sample. A microscope objective, a polarizer and a CCD-camera
  are used to detect the intensity in the target focus, a few millimeters behind the sample.
  }\label{fig:setup}
\end{figure}

\noindent The different algorithms were tested experimentally using the setup shown in Fig.
\ref{fig:setup}. In our case the scattering medium is a 10 $\mu$m thick layer of rutile TiO$_2$ pigment
\cite{Kop1997} with a mean free path of $0.55\pm0.1$ $\mu$m, determined by measuring the total
transmission at a wavelength of 632.8 nm. This sample is illuminated by a 632.8 nm HeNe laser. The laser
beam is expanded and spatially modulated by a Holoeye R-2500 liquid crystal light modulator (LCD)
operating in phase-mostly modulation mode \cite{Davis2002}. The shaped beam is focused on the sample using
a 63x objective with a numerical aperture (NA) of 0.85. A 20x objective (NA=0.5) images a point that is
approximately $3.5$ mm behind the sample onto a 12-bit CCD camera (Allied Vision Technologies Dolphin
F-145B). This point is the target area where we want the light to focus. A computer integrates the
intensity in a circular area with a radius of 20 pixels (corresponding to 129 $\mu$m in the focal plane of
the objective). This target area is smaller than a typical speckle spot. Using this signal as feedback,
the computer programs the phase modulator using one of the algorithms described above.

\begin{figure}
  \includegraphics[width=7.5cm]{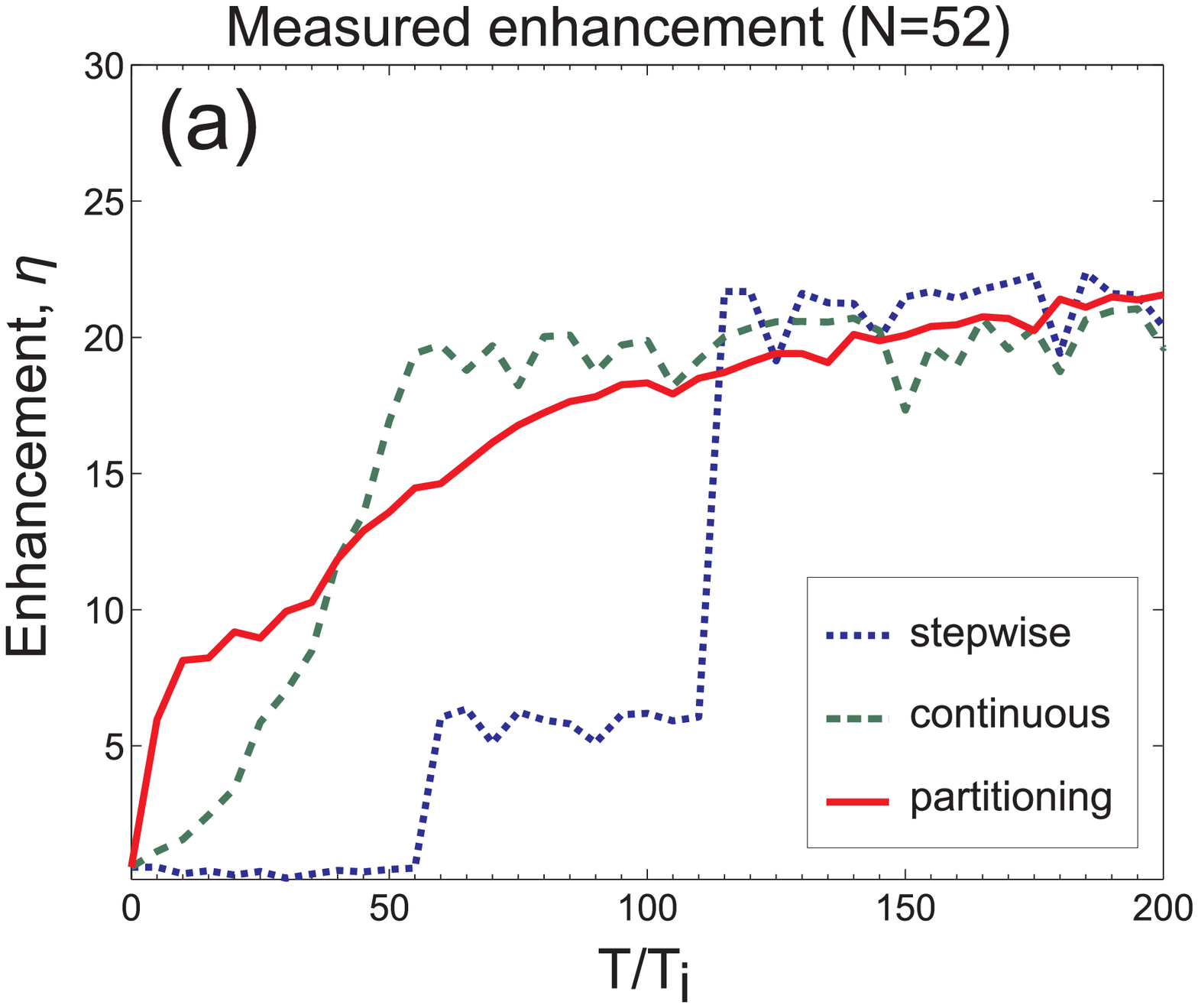}
  \\[20pt]
  \includegraphics[width=7.5cm]{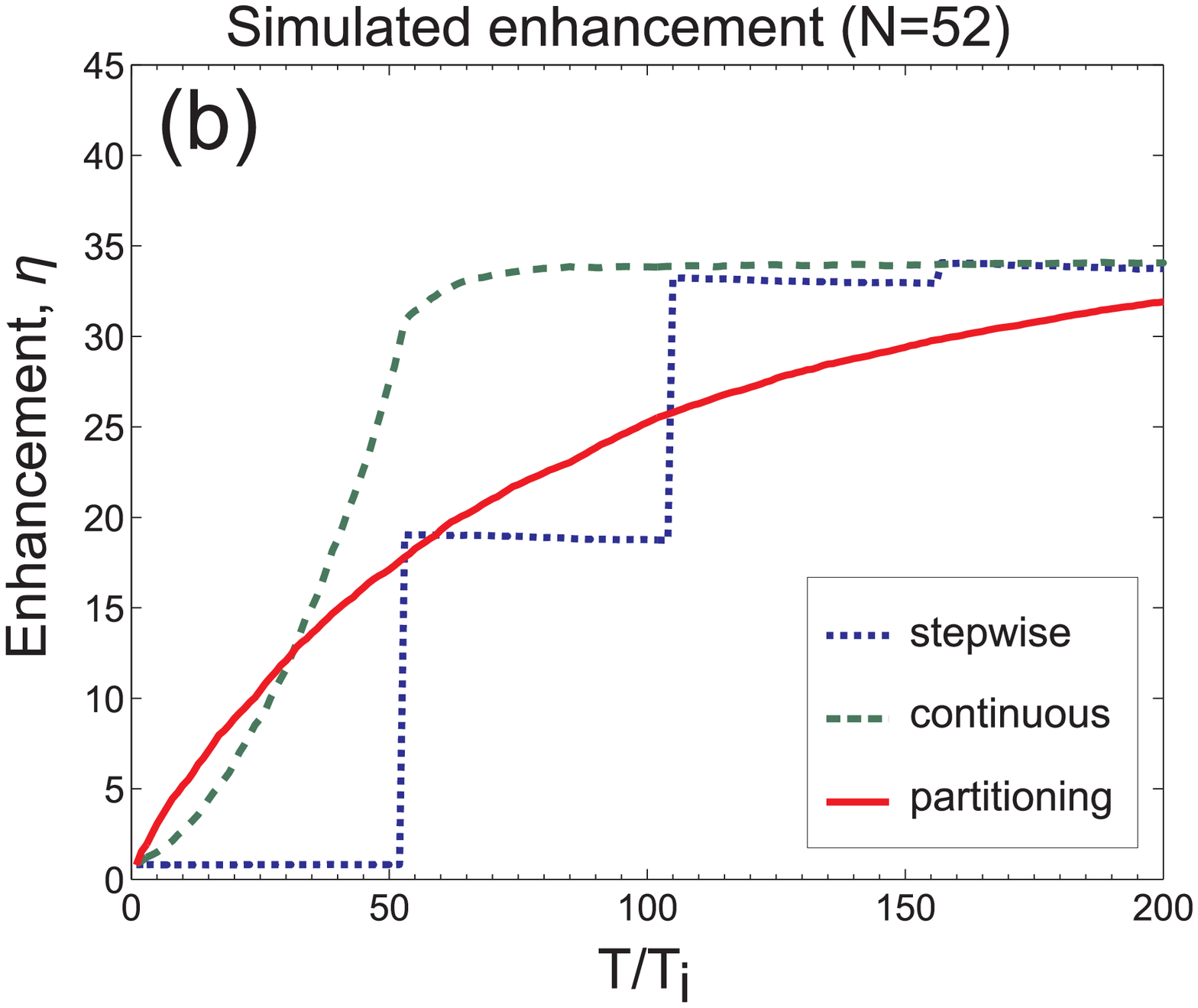}
  \caption{(a) Typical runs of the stepwise sequential algorithm (dotted line), the continuous sequential algorithm (dashed line)
  and the partitioning algorithm (solid line). All algorithms were run with $N=52$.
  The sequential algorithms were repeated four times. (b) Simulation results for $N=52$ averaged over 64 runs. The
  simulation captures the main features of the three algorithms, but predicts a higher maximum enhancement. } \label{fig:low-N}
\end{figure}

We first run the three different algorithms with $N=52$. Since $T_p/T_i \gg 52$, we do not expect to see
decoherence effects. In total, 208 iterations were performed, which means that the sequential algorithms
ran four times consecutively. The results of the optimization procedures is shown in Fig.
\ref{fig:low-N}a. Although the three algorithms reach the same final enhancement of intensity, there are
significant differences between the algorithms. The enhancement for the stepwise sequential algorithm
increases in discrete steps because the phase modulator is only reprogrammed every $N$ iterations. During
the first $N$ iterations, the target signal is low and the algorithm suffers from noise. The saturation
enhancement is reached after the second update (after $2N$ iterations). The continuous sequential
algorithm and the partitioning algorithm both start updating the wavefront immediately and, therefore,
have a higher initial increase of the signal. The continuous sequential algorithm is the first algorithm
to reach the saturation enhancement (after $N$ iterations). The partitioning algorithm has the fastest
initial increase in the target signal. It is, however, the last algorithm to reach the saturation
enhancement since the final convergence is very slow.

\begin{figure}
  \includegraphics[width=7.5cm]{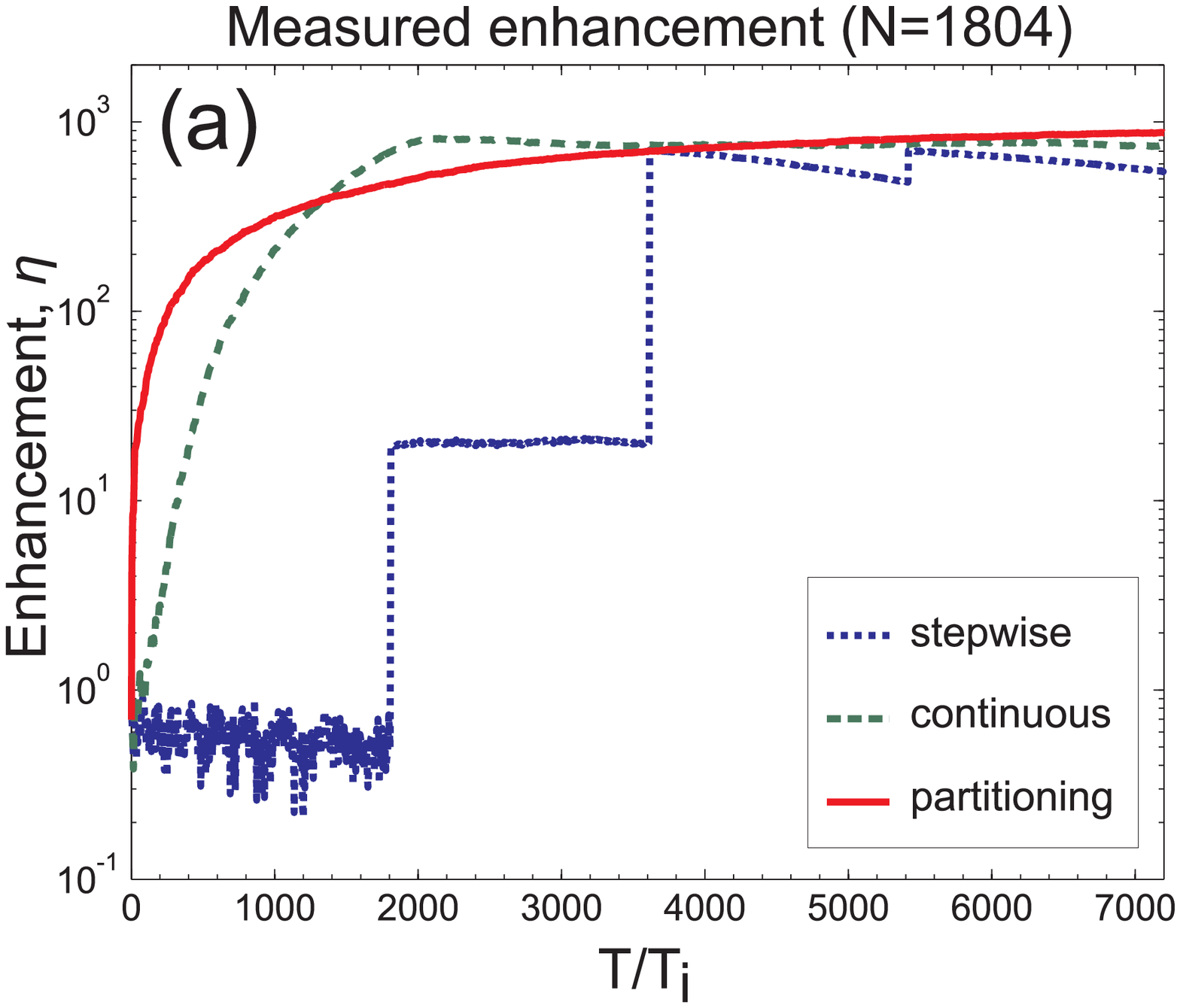}
  \\[20pt]
  \includegraphics[width=7.5cm]{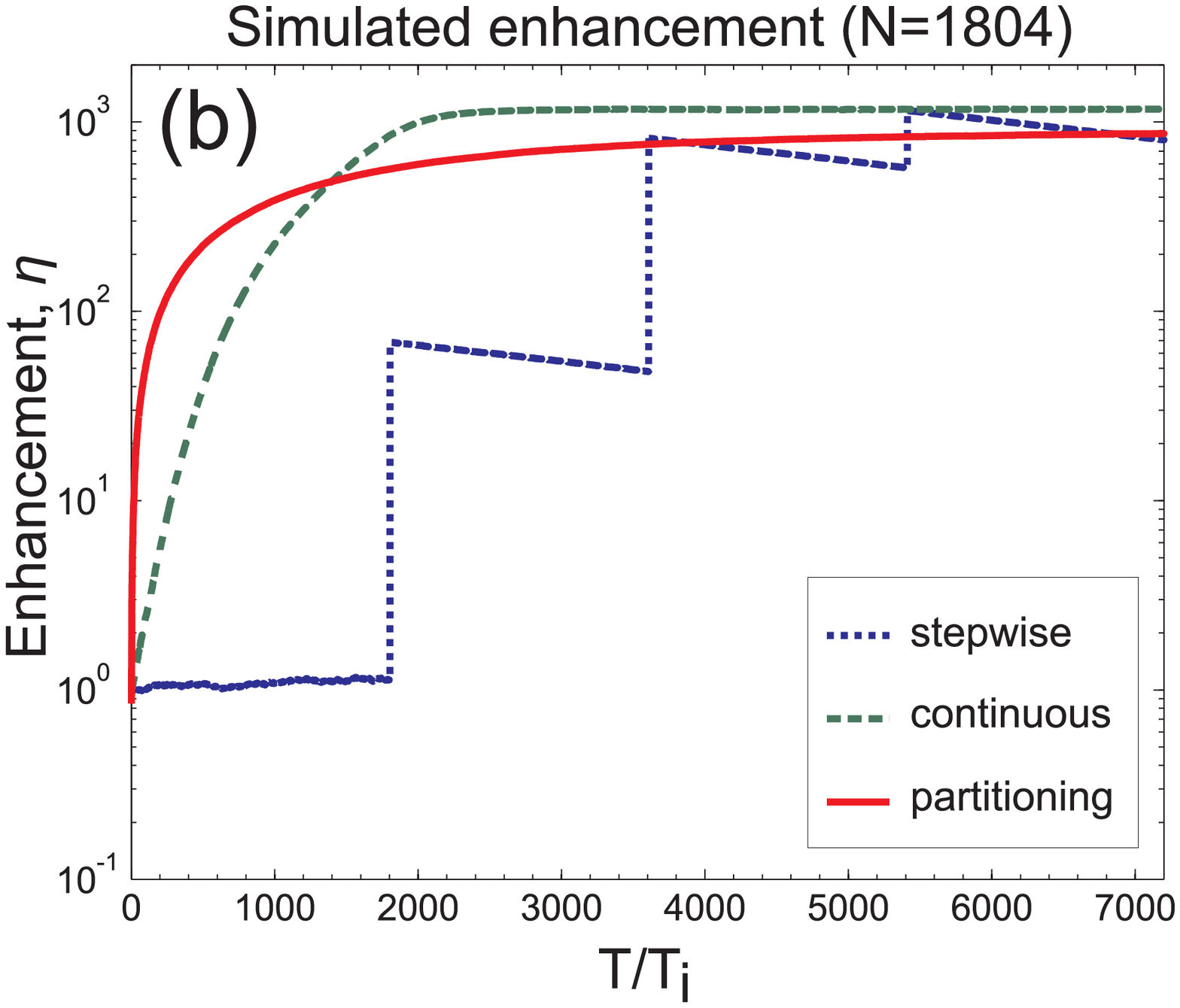}
  \caption{(a) Typical runs of stepwise sequential algorithm (dotted line), the continuous sequential algorithm (dashed line)
  and the partitioning algorithm (solid line). All algorithms were run with $N=1804$. The enhancement are plotted on a logarithmic scale.
  (b) Simulation results for $N=1804$ averaged over 64 runs.
  }\label{fig:high-N}
\end{figure}

When the number of segments in the wavefront is increased, we expect to find a higher target intensity.
Figure \ref{fig:high-N}a shows the experimental results for $N=1804$ on a logarithmic scale. The final
intensity enhancement is approximately 40 times higher than in Fig. \ref{fig:low-N}a. A further difference
is that in this situation the effects of decoherence are no longer negligible. This effect is most clearly
visible with the stepwise sequential algorithm. The phase modulator is updated after each $N$ iterations
and between the updates the intensity decays exponentially with a $1/e$ decay of about $T_p/T_i=5000$
iterations.

The convergence behavior of the three algorithms is similar to the
experiment shown in Fig. \ref{fig:low-N}. The partitioning algorithm
clearly causes a higher signal enhancement during the first 1000
iterations. The initial increase in the enhancement is linear with a
slope of 0.37. Initially, this linear increase is far superior to
the quadratic increase obtained with the continuous sequential
algorithm.

We conclude that both new algorithms are valuable improvements over the original stepwise sequential
algorithm. These algorithms are far less sensitive to noise and the target signal is kept at a constant
value even in the presence of decoherence. The partitioning algorithm has the fastest initial increase and
therefore will recover from disturbances most rapidly.

\section{Simulations}
\noindent In order to obtain a better understanding of the effects of noise and fluctuations on the
performance of the different algorithms we perform numerical simulations. The disordered medium is
represented by a transmission matrix with elements drawn from a circular Gaussian distribution (more
details on the matrix representation can be found below). Decoherence is modelled by adding a small
perturbation to the transmission matrix after every measurement. Finally, measurement noise is included by
adding a random value to the simulated detector signal.

Figures \ref{fig:low-N}b and \ref{fig:high-N}b show the simulated enhancement for a system with $T_p/T_i
=5000$. Every iteration, ten measurements are performed for phase delays between $0$ and $2 \pi$. To each
of these measurements, Gaussian noise with a standard deviation of $0.3 I_0$ was added. The magnitude of
the noise is comparable to experimental observations. The three different algorithms were run with $N=52$
and $N=1804$ to simulate the experiments shown in Fig. \ref{fig:low-N}a and Fig. \ref{fig:high-N}a.

The simulations are in good qualitative agreement with the
experimental data. The result for the stepwise sequential algorithm
shows that the effects of noise and decoherence are simulated
realistically. Furthermore, the initial signal increase and the long
time convergence behavior correspond to the measured results. The
only significant difference is the 20\% to 50\% higher enhancement
reached in the simulations. A possible explanation for this
difference is the residual amplitude cross-modulation in our phase
modulator. Due to this cross-modulation, the optimal wavefront
cannot be generated exactly. Furthermore, the amplitude modulation
decreases the accuracy of the measurement of the optimal phase. The
partitioning algorithm is less sensitive to this last effect since
the cross-modulation is averaged over many segments with different
phases. Since the simulations capture the overall behavior of the
algorithms very well, we can use them to extrapolate to situations
with a lot of noise and strong decoherence or, on the other hand, to
perfectly stable systems.

\section{Analytical expressions for the enhancement}
\noindent In this section, we analyze the performance of the
algorithms with analytical theory and compare these results to the
simulations. We describe scattering in the sample with the
transmission matrix elements, $t_{mn}$. This matrix couples the
fields of the incident light and the transmitted light.
\begin{equation}
E_m = \sum_n^N t_{mn} A_n e^{i\phi_n},
\end{equation}
where the $\phi_n$ is the phase of the $n$th segment of the phase modulator. Assuming that the modulator
is illuminated homogeneously, all incoming channels carry the same intensity. We write $A_n = 1/\sqrt{N}$
to normalize the total incident intensity. Elements $E_1, E_2, \ldots$ correspond to single scattering
channels of the transmitted light. Since we are interested in focusing light to a single spot, we need to
consider only a single transmission channel, $E_m$. The intensity transmitted into channel $m$ is given by
\begin{equation}
|E_m|^2 = \frac1N\left|\sum_n^N t_{mn}
e^{i\phi_n}\right|^2\label{eq:intensity-base}.
\end{equation}
Regardless of the values of the elements $t_{mn}$ of the transmission matrix, the intensity $|E_m|^2$ has
its global maximum when the phase modulator exactly compensates the phase retardation in the sample for
each segment, i.e. $\phi_n=-\arg(t_{mn})$. The target intensities before optimization ($I_0$) and after an
ideal optimization ($I_\text{max}$) are given by
\begin{equation}
I_0 = \frac1N\left|\sum_n^N t_{mn}\right|^2 \label{eq:intensity-before},
\end{equation}
and
\begin{equation}
I_\text{max} = \frac1N \left(\sum_n^N \left| t_{mn} \right|\right)^2\label{eq:intensity-after}.
\end{equation}

For a disordered medium the elements of $t_{mn}$ are independent and have a Gaussian distribution
\cite{Goodman2000,Garcia1989,Webster2004,Beenakker1997}. Rewriting Eq. \eqref{eq:intensity-after} gives
\begin{align}
\avg{I_\text{max}} &= \avg{\frac1N\sum^N_{n, k\neq n}
|t_{mn}||t_{mk}| + \frac1N \sum_n^N |t_{mn}|^2}, \label{eq:intensity-after-intermediate}\\
& = \avg{I_0}\left[(N-1)\frac{\pi}{4}+1\right], \label{eq:intensity-after-expanded}
\end{align}
where the angled brackets denote ensemble averaging over disorder. Eq. \eqref{eq:intensity-after-expanded}
predicts that the expected maximum enhancement for an ideally stable, noise free system linearly depends
on the number of segments $N$. For $N \gg 1$, we have $\eta \approx \pi N /4$.

\subsection{Performance in fluctuating environments}
\noindent In reality, the sample will not be completely stable. Whether this instability is due to a drift
of the sample position, movement of the scatterers, changing humidity or any other cause, the transmission
matrix will fluctuate over time. In the simulations, we modelled decoherence by repeatedly adding a small
perturbation to each of the matrix elements.
\begin{equation}
t_{mn}\rightarrow \frac{1}{\sqrt{1+\delta^2}}(t_{mn} +
\xi)\label{eq:decoherence-principle},
\end{equation}
where $\xi$ is drawn from a complex Gaussian distribution with mean 0 and standard deviation $\delta$. The
prefactor normalizes the transformation so that $\avg{|t|^2}$ remains constant. By substituting the
continuous limit of Eq. \eqref{eq:decoherence-principle} in Eq. \eqref{eq:intensity-after-intermediate},
we find an analytic expression for the effect of decoherence,

\begin{equation}
\avg{I_N} = \avg{I_0} \left[\frac{\pi}{4N}\left(\sum^N_n
e^{-T_n\delta^2/(2T_i)}\right)^2+O(1)\right],\label{eq:intensity-decoherence}
\end{equation}
where $T_n$ is the time that has past since the phase of segment $n$ was measured. This simple model
explains the exponential decay of the intensity that was observed in the measurements (see
Fig.\ref{fig:high-N}a) and predicts a decay time of $T_p= T_i/\delta^2$.

We now calculate the maximum enhancement that can be reached with the continuous sequential algorithm in
the presence of decoherence. Because the phases of the segments are measured sequentially, at any given
time the values for $T_n$ are equally spaced between $1$ and $N$. From Eq.
\eqref{eq:intensity-decoherence} we find a maximum intensity enhancement of
\begin{equation}
\eta_N\equiv\frac{\avg{I_N}}{\avg{I_0}}=
\frac{\pi}{4N}\left(\frac{1-e^{-N
T_i/(2T_p)}}{e^{T_i/(2T_p)}-1}\right)^2+O(1).\label{eq:intensity-alg12}
\end{equation}

The maximum enhancement for both sequential algorithms is the same.
However, since the stepwise algorithm only updates the projected
wavefront after $N$ iterations, the enhancement decreases
exponentially between updates.

\begin{figure}
  \includegraphics[width=7.5cm]{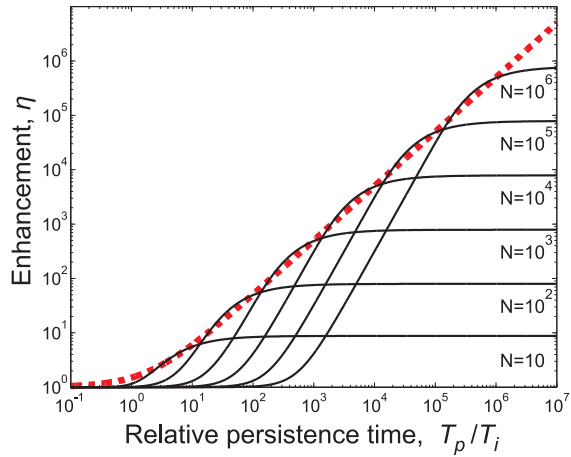}\\
  \caption{Theoretical maximum enhancement as a function of coherence time for different algorithms. The solid lines represent the maximum enhancement
  that can be obtained using the sequential algorithms. The enhancement depends on the number of segments used in the
  algorithm. The dashed line shows the enhancement for the partitioning algorithm where $N\gg T_p/T_i$.}\label{fig:algorithm-comparison-decoherence}
\end{figure}

In Fig. \ref{fig:algorithm-comparison-decoherence} the enhancement
for different values of $N$ is plotted versus $T_p/T_i$. When the
persistence time is large ($T_p/T_i \gg N$), decoherence effects do
not play a role and the enhancement linearly depends on $N$ as was
seen in Eq. \eqref{eq:intensity-after-expanded}. For $T_p/T_i < N$,
however, the enhancement decreases because the speckle pattern
decorrelates before all iterations are performed and the enhancement
drops to zero. As a consequence, the sequential algorithms only
perform optimal when $N$ is adjusted to $T_p$. When $T_p$ is known
a-priori, this optimum for $N$ can be found by maximizing Eq.
\eqref{eq:intensity-alg12}. We find
\begin{equation}
N_\text{opt}=W T_p/T_i\label{eq:N-optimal},
\end{equation}
where $W\approx 2.51$ is the solution of $\exp(W/2)=1+W$. The maximal enhancement achievable with
sequential algorithms follows by substituting Eq. \eqref{eq:N-optimal} into Eq. \eqref{eq:intensity-alg12}
and equals $\eta_\text{opt} = 0.640 T_p/T_i$.

With the partitioning algorithm $\eta$ increases by $1/2$ each iteration of the algorithm (see appendix
\ref{sec:alg3}). As long as $N \gg T_p/T_i$, the enhancement saturates at $\eta = T_p/(2 T_i) + 1$, when
the increase is exactly cancelled by the effect of decoherence. The most important difference with the
sequential algorithms, is that the enhancement reached with the partitioning algorithm does not depend on
$N$. In Fig. \ref{fig:algorithm-comparison-decoherence} it can be seen that the partitioning algorithm
outperforms the sequential algorithms for almost all combinations of $T_p$ and $N$. The sequential
algorithm only give a slightly higher enhancement when they are fine-tuned for a known persistence time
($N = 2.51 T_p/T_i$). In most situations, $T_p$ is not known a-priory or varies over time and the
partitioning algorithm will be preferable.

\begin{figure}
  \includegraphics[width=7.5cm]{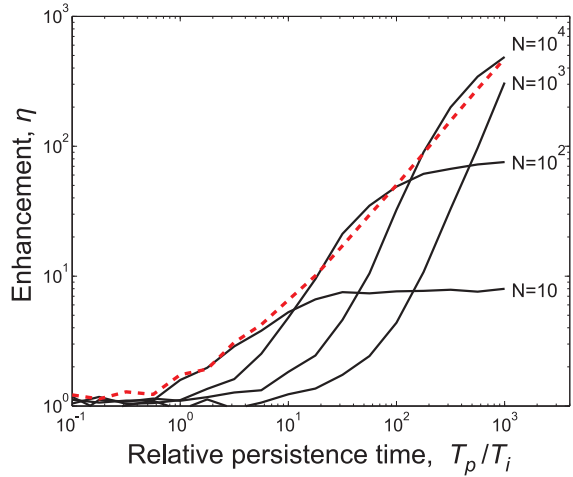}\\
  \caption{Simulated effect of decoherence on the sequential algorithms (solid line) and on the partitioning algorithm (dashed
  line). The simulations are averaged over 25 runs. Only when $N \approx 2.51 T_p/T_i$, the sequential algorithms perform slightly better than the partitioning
  algorithm.}\label{fig:algorithm-decoherence-simulation}
\end{figure}

Our analytical results for all three algorithms are supported by numerical simulations (see Fig.
\ref{fig:algorithm-decoherence-simulation}). The simulations exactly reproduce the theoretical curves
shown in Fig. \ref{fig:algorithm-comparison-decoherence}. For the simulations we used $N=4096$, a number
that can easily be reached with a LCD phase modulator. Again, the partitioning algorithm can be seen to
have good overall performance, whereas the sequential algorithms only work well for certain combinations
of $N$ and $T_p$.

In conclusion, the maximum enhancement that can be reached linearly depends on the sample's persistence
time. For the sequential algorithms $\eta = 0.64 T_p/T_i$, but only when $N$ is precisely adjusted to
$T_p$. The partitioning algorithm has $\eta = 0.5 T_p/T_i$, as long as $N$ is large enough. Using these
analytical relations, the performance of each of the three algorithms in different experimental situations
can easily be estimated.

\section{Effect of Noise}
\begin{table}
\begin{tabular}{|l|c|c|c|}
\hline
          & stepwise   & continuous & partitioning\\
          & sequential & sequential &\\
\hline signal & $2 I_0 \sqrt{1/N}$ & $2 I_0 \sqrt{\eta/N}$ & $\eta I_0$\\
bias & $I_0$ & $\eta I_0$ & $\eta I_0$\\
relative shot noise SNR & $2\sqrt{I_0/N}$ & 2$\sqrt{I_0/N}$ & $\sqrt{\eta I_0}$\\
rms phase correction & $\sqrt{3}\pi$ & $\sqrt{3}\pi$ & $\sqrt{2/\eta}$\\
\hline
\end{tabular}
\caption{Signal and noise characteristics of the three algorithms. The rms phase correction is a measure
for the required sensitivity.}\label{tab:noise}
\end{table}

\noindent Measurement noise affects the measured phases. Noise induced errors in the phases lead to a
reduction of the enhancement, $\eta$. We will now compare the signal-to-noise ratio (SNR) of the three
different algorithms. In a single iteration of an algorithm, the phase of one or more segments is varied,
while the phase of the other segments is kept constant. The intensity at the detector equals
\begin{equation}
I(\Phi) = I_A + I_B + 2\sqrt{I_A I_B} \cos{(\Phi - \Phi_0)},\label{eq:signal}
\end{equation}
where $I_B$ is the intensity at the target originating from the modulated segments, $I_A$ is the target
intensity caused by light coming from the other segments, $\Phi$ is the phase that is varied and $\Phi_0$
is the unknown optimal value for the phase. The last term in Eq. \eqref{eq:signal} is the signal that is
relevant for measuring $\Phi_0$. There first two terms constitute a constant bias.

Table \ref{tab:noise} lists the magnitudes of the signal and the bias for each of the three algorithms. If
the detection system is photon shot noise limited, the noise is proportional to the square root of the
bias. When, on the other hand, constant noise sources such as readout noise or thermal noise are dominant,
the SNR is directly proportional to the signal magnitude. Table \ref{tab:noise} also shows the root mean
square (rms) phase correction that is applied during each iteration of the corresponding algorithm. The
rms phase correction is a measure for the required accuracy of the measurements.

With the stepwise sequential algorithm, $I_A \approx I_0$ and on average $I_B = I_0/N$. Since the initial
diffuse transmission $I_0$ is low and $N$ can be very high, the SNR is low. The continuous sequential
algorithm has a higher SNR since the overall intensity at the detector increases while the algorithm
progresses and $I_A \approx \eta I_0$. Assuming the dominant noise source is constant, the SNR will
increase as the enhancement becomes higher. Therefore, the algorithm can be accelerated by decreasing the
integration time of the camera as the algorithm advances. In case the detection system is photon shot
noise limited, the SNR remains constant during the optimization since both the signal and the shot noise
scale as $\sqrt{\eta I_0}$.

The highest SNR is achieved with the partitioning algorithm. Since we always change the phase of half of
the segments, $I_A \approx I_B \approx \eta I_0/2$, resulting in a maximal signal. Unlike the sequential
algorithms, the SNR does not depend on $N$. Therefore the number of segments can be increased without
suffering from noise. Like with the continuous sequential algorithm, the integration time can be adjusted
dynamically to optimize the speed/SNR tradeoff. Although the partitioning algorithm has the highest SNR,
the magnitude of the phase corrections decreases as the algorithm progresses. The required accuracy in
measuring $\Phi_0$ increases at the same pace as the SNR increases.

The partitioning algorithm is very sensitive to measurement errors since, when the measured $\Phi_0$ has
an error, half of the segments will be programmed with the wrong phase. In the extreme case where the
error equals $\pi$ the enhancement completely disappears in a single iteration. A simple and effective
solution to this problem is to keep the previous configuration of the phase modulator in memory. When an
optimization step causes the signal to decrease, the algorithm can revert to the saved configuration.

\section{Conclusion}
\noindent Three different algorithms for inverting wave diffusion were presented. The algorithms were
compared experimentally, with numerical simulations and using analytical theory. We found good agreement
between experimental data, simulations and theory. Moreover, the simulations and theory can be used to
predict the performance in different experimental situations.

The effectiveness of the algorithms was quantified by the enhancement. It was seen that the enhancement
depends on the number of segments $N$ and the relative persistence time $T_p/T_i$. For the sequential
algorithms to have optimal performance, it is required to adjust $N$ to match $T_p$. This means that these
algorithms need a-priori knowledge of the system. The partitioning algorithm does not need this knowledge
and always performs close to optimal. Moreover, the algorithm causes the enhancement to increase the most
rapidly of the three investigated methods. All in all, this algorithm is a good candidate for applying
inverse diffusion in instable scattering media such as living tissue. In the future, learning algorithms
(see e.g. \cite{Goldberg1989,Judson1992}) might be developed to further improve the performance of inverse
diffusion, for instance by dynamically balancing the trade-off between signal to noise ratio and
measurement speed.

The maximum enhancement linearly depends on the number of measurements that can be performed before the
speckle pattern decorrelates ($T_p/T_i$). The faster the measurements, the higher the enhancement. In our
current system, the speed is limited by the response time of the LCD. Fast micro mechanical phase
modulators have a mechanical response time of about 10 $\mu$s (see e.g. \cite{Hacker2003}), which allows a
$10^4$ times faster operation than with our current system. In perfused tissue, a typical decorrelation
timescale is 10 ms \cite{Briers1995}, which means that an enhancement of about 50 should be possible with
currently available technology.

\section{Acknowledgements}
\noindent We thank Prof. Ad Lagendijk and Prof. Willem Vos for support and valuable discussions. This work
is part of the research program of the ``Stichting voor Fundamenteel Onderzoek der Materie (FOM)", which
is financially supported by the ``Nederlandse Organisatie voor Wetenschappelijk Onderzoek (NWO)".

\appendix
\section{Calculation of the performance of the partitioning algorithm}\label{sec:alg3}
\noindent In this appendix we calculate the development of the enhancement of the partitioning algorithm
under ideal conditions. During one iteration of the partitioning algorithm, the phase modulator is
randomly split into two groups ($A$ and $B$), each containing half of the segments. The relative phase
($\Phi$) of group $B$ is cycled from 0 to $2\pi$. During this cycle, the target intensity is given by
\begin{equation}
I(\Phi) = \left| E_{mA} + E_{mB}e^{i\Phi}\right|^2,\label{eq:random-intensity-before}
\end{equation}
where $E_{mA}$ is the contribution of the segments in group $A$ to the target field
\begin{equation}
E_{mA} = \sum_{n\in A} \sqrt{\frac{\avg{I_0}}{N}}\xi_{mn},\label{eq:partitioning-A-xi}
\end{equation}
with
\begin{equation}
\xi_{mn} \equiv \sqrt{\frac{1}{\avg{I_0}}} t_{mn} e^{i\phi_n},
\end{equation}
and similar for $E_{mB}$. The coefficients $\xi_{mn}$ are initially random and distributed according to a
normalized circular Gaussian distribution, meaning that $\avg{\xi}=0$ and
$\avg{(\Re{\xi})^2}=\avg{(\Im{\xi})^2}=1/2$. As the algorithm proceeds, the phases $\phi_n$ are adjusted
and the distribution gradually changes to a Rayleigh distribution when a high enhancement is reached. The
average value $\avg{\xi}$ increases from $0$ to $\sqrt{\pi/4}$ as all contributions are aligned to be in
phase. At any moment during the optimization, $\avg{\xi} = \sqrt{(\eta-1)/(N-1)}$ and $\avg{|\xi|^2}=1$.

\begin{figure}
\includegraphics[width=4cm]{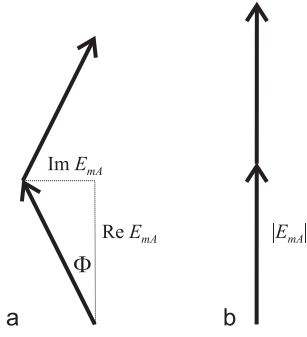}
\caption{Complex plane representation of the partitioning algorithm. a) Before the iteration the
contributions from $A$ and $B$ are not exactly in phase. b) After the iteration, the contributions are
aligned and the resulting intensity is higher.}\label{fig:vectors}
\end{figure}

Figure \ref{fig:vectors} gives a graphical representation of a single iteration. Before the iteration,
$E_{mA}$ and $E_{mB}$ have a different phase. Without loss of generality, we choose the phase of
$(E_{mA}+E_{mB})$ to be $0$. The intensity before the iteration is given by
\begin{equation}
I_\text{before} = \left( \Re E_{mA} + \Re E_{mB} \right)^2.
\end{equation}
After the iteration, $\Phi$ is set to the value that caused the highest target intensity, which means that
$E_{mA}$ and $E_{mB}$ are now in phase. The target intensity then equals
\begin{equation}
I_\text{after} = \left( |E_{mA}| + |E_{mB}| \right)^2,\label{eq:random-intensity-after}
\end{equation}
which is higher than or equal to before the iteration.

We now calculate the average intensity gained in a single iteration. We consider the regime where already
a few iterations have been done ($\eta \gg 1$). In this regime, we can approximate
\begin{align}
|E_{mA}| &= \sqrt{(\Re E_{mA}) ^2 + (\Im E_{mA})^2}\\
&\approx \Re E_{mA} + \frac{(\Im E_{mA})^2}{2\Re E_{mA}}.
\end{align}
Using this result in Eq. \eqref{eq:random-intensity-after} gives
\begin{multline}
I_\text{after} = \left( \Re E_{mA} + \Re E_{mB}| \right)^2 + (\Im E_{mA})^2 + (\Im E_{mB})^2 +\\
 + \frac{\Re{E_{mA}}}{\Re{E_{mB}}}(\Im E_{mB})^2 + \frac{\Re{E_{mB}}}{\Re{E_{mA}}}(\Im E_{mA})^2 +\\
+ \frac{(\Im E_{mA})^4}{4(\Re E_{mA})^2}+ \frac{(\Im E_{mB})^4}{4(\Re E_{mB})^2} + \frac{(\Im E_{mA} \Im
E_{mB})^2}{2\Re E_{mA}\Re E_{mB}}
\end{multline}
where the terms on the last line can be neglected. If $N\gg1$, \\ $\Re{E_{mB}}/\Re{E_{mA}}\approx 1$. The
intensity gain for the iteration $\Delta I \equiv I_\text{after}-I_\text{before}$ is now found to be
\begin{equation}
\Delta I = 2(\Im E_{mA})^2 + 2(\Im E_{mB})^2
\end{equation}\label{eq:random-intensity-increase}
We are primarily interested in the regime $N\gg\eta\gg1$), where the algorithm picks up the main part of
the final enhancement. In this regime, $\avg{\xi} \ll 1$ and the probability distribution of $\xi$ is
still close to the original Gaussian distribution. Therefore, $\avg{(\Im{\xi})^2}\approx 1/2$ and it
follows from Eq. \eqref{eq:partitioning-A-xi} that
\begin{equation}
\Delta I = \frac12\avg{I_0}
\end{equation}
Therefore, we expect the intensity enhancement $\eta$ to increase with $1/2$ after each iteration of the
algorithm. With this information, we also calculate the typical phase adjustment that is performed in each
iteration. From Fig. \ref{fig:vectors} it follows that the root mean square phase adjustment equals
\begin{equation}
\Phi_{\text{rms}} \equiv \sqrt{\avg{\Phi}} = \sqrt{\avg{\frac{(\Im E_{mA})^2}{(\Re E_{mA})^2}}} =
\sqrt{\frac{2}{\eta}}
\end{equation}
When $\eta$ approaches its maximum, all contributions are almost completely in phase and
$\avg{(\Im{\xi})^2}$ vanishes. In this regime, the algorithm becomes less and less effective, as was seen
in simulations and experiments (see Fig. \ref{fig:high-N}).


\begin{thebibliography}{10}
\expandafter\ifx\csname url\endcsname\relax
  \def\url#1{\texttt{#1}}\fi
\expandafter\ifx\csname urlprefix\endcsname\relax\def\urlprefix{URL }\fi

\bibitem{Milne1921}
E.~A. Milne, Radiative equilibrium in the outer layers of a star, Monthly Not.
  Roy. Astron. Soc. 81 (1921) 361--375.

\bibitem{Chandrasekhar1960}
S.~Chandrasekhar, Radiative Transfer, Dover Publications, Inc., New York, 1960.

\bibitem{Ishimaru1978}
A.~Ishimaru, Limitation on image resolution imposed by a random medium, Appl.
  Opt. 17 (1978) 348--352.

\bibitem{Vellekoop2007}
I.~M. Vellekoop, A.~P. Mosk, Focusing coherent light through opaque strongly
  scattering media, Optics Lett. 32~(16) (2007) 2309--2311.

\bibitem{Tyson1998}
R.~K. Tyson, Principles of Adaptive Optics, 2nd Edition, Academic Press, 1998.

\bibitem{Fink1999}
M.~Fink, D.~Cassereau, A.~Derode, C.~Prada, P.~Roux, M.~Tanter, J.-L. Thomas,
  F.~Wu, Time-reversed acoustics, Rep. Prog. Phys. 63 (1999) 1933--1995.

\bibitem{Lobkis2001}
O.~I. Lobkis, R.~L. Weaver, On the emergence of the green's function in the
  correlations of a diffuse field, J. Ac. Soc. Am. 110~(6) (2001) 3011--3017.

\bibitem{Borcea2005}
L.~Borcea, G.~Papanicolaou, C.~Tsogka, Interferometric array imaging in
  clutter, Inv. Prob. 21 (2005) 1419--1460.

\bibitem{Lerosey2007}
G.~Lerosey, J.~de~Rosny, A.~Tourin, M.~Fink, Focusing beyond the diffraction
  limit with far-field time reversal, Science 315 (2007) 1120--1122.

\bibitem{Foschini1996}
G.~J. Foschini, Layered space-time architecture for wireless communication in a
  fading environment when using multi-element antennas, Bell Labs Technical
  Journal 1 (1996) 41--59.

\bibitem{Shapiro2005}
N.~M. Shapiro, M.~Campillo, L.~Stehly, M.~H. Ritzwoller, High-resolution
  surface-wave tomography from ambient seismic noise, Science 307 (2005) 1615.

\bibitem{Kuperman1998}
W.~A. Kuperman, W.~S. Hodgkiss, H.~C. Song, T.~Akal, C.~Ferla, D.~R. Jackson,
  Phase conjugation in the ocean: Experimental demonstration of an acoustic
  time-reversal mirror, J. Acoust. Soc. Am. 103 (1998) 25--40.

\bibitem{Stockman2002}
M.~I. Stockman, S.~V. Faleev, D.~J. Bergman, Coherent control of femtosecond
  energy localization in nanosystems, Phys. Rev. Lett. 88~(6) (2002) 067402.

\bibitem{Herek2002}
J.~L. Herek, W.~Wohlleben, R.~J. Cogdell, D.~Zeidler, M.~Motzkus, Quantum
  control of energy flow in light harvesting., Nature 417 (2002) 533--535.

\bibitem{Pappu2002}
R.~Pappu, B.~Recht, J.~Taylor, N.~Gershenfeld, Physical one-way functions,
  Science 297 (2002) 2026--2030.

\bibitem{Briers1995}
J.~D. Briers, S.~Webster, Quasi real-time digital version of single-exposure
  speckle photography for full-field monitoring of velocity or flow fields,
  Opt. Commun. 116~(1) (1995) 36--42.

\bibitem{Kop1997}
R.~H.~J. Kop, P.~de~Vries, R.~Sprik, A.~Lagendijk, Observation of anomalous
  transport of strongly multiple scattered light in thin disordered slabs,
  Phys. Rev. Lett. 79~(22) (1997) 4369--4372.

\bibitem{Davis2002}
J.~A. Davis, J.~Nicol{\'a}s, A.~M{\'a}rquez, Phasor analysis of eigenvectors
  generated in liquid-crystal displays, Appl. Opt. 41~(22) (2002) 4579--4584.

\bibitem{Goodman2000}
J.~W. Goodman, Statistical optics, Wiley, New York, 2000.

\bibitem{Garcia1989}
N.~Garcia, A.~Z. Genack, Crossover to strong intensity correlation for
  microwave radiation in random media, Phys. Rev. Lett. 63 (1989) 1678--1681.

\bibitem{Webster2004}
M.~A. Webster, T.~D. Gerke, A.~M. Weiner, K.~J. Webb, Spectral and temporal
  speckle field measurements of a random medium, Opt. Lett. 29~(13) (2004)
  1491.

\bibitem{Beenakker1997}
C.~W.~J. Beenakker, Random-matrix theory of quantum transport, Rev. Mod. Phys.
  69 (1997) 731--808.

\bibitem{Goldberg1989}
D.~E. Goldberg, Genetic algorithms in search, optimization \& machine learning,
  Addison-Wesley, Reading, MA, 1989.

\bibitem{Judson1992}
R.~S. Judson, H.~Rabitz, Teaching lasers to control molecules, Phys. Rev. Lett.
  68 (1992) 1500–--1503.

\bibitem{Hacker2003}
M.~Hacker, G.~Stobrawa, R.~Sauerbrey, T.~Buckup, M.~Motzkus, M.~Wildenhain,
  A.~Gehner, Micromirror slm for femtosecond pulse shaping in the ultraviolet,
  Appl. Phys. B 76 (2003) 711.

\end{thebibliography}

\end{document}